\documentclass[pre,twocolumn,amsmath,amssymb,floatfix,superscriptaddress]{revtex4-1}
\pdfoutput=1
\usepackage{graphicx}
\usepackage{dcolumn}
\usepackage{bm}

\usepackage{subfigure}

\graphicspath{{images/}{plots/}}

\newcommand{\avg}[1]{\ensuremath{\left< #1 \right>}}

\newcommand{\brac}[1]{\ensuremath{\left(#1\right)}}

\begin{document}

    \title{Replica Symmetry and Replica Symmetry Breaking for the Traveling Salesperson Problem}
    \author{Hendrik Schawe}
    \email{hendrik.schawe@uni-oldenburg.de}
    \affiliation{Institut f\"ur Physik, Universit\"at Oldenburg, 26111 Oldenburg, Germany}
    \author{Jitesh Kumar Jha}
    \email{jiteshjha96@gmail.com}
    \affiliation{Institut f\"ur Physik, Universit\"at Oldenburg, 26111 Oldenburg, Germany}
    \affiliation{Manipal Institute of Technology, 576104 Karnataka, India}
    \author{Alexander K. Hartmann}
    \email{a.hartmann@uni-oldenburg.de}
    \affiliation{Institut f\"ur Physik, Universit\"at Oldenburg, 26111 Oldenburg, Germany}
    \date{\today}

    \begin{abstract}
        We study the energy landscape of the Traveling Salesperson problem (TSP)
        using exact ground states and a novel linear programming approach to
        generate excited states with closely defined properties.
        We look at four different ensembles, notably the classic finite dimensional
        Euclidean TSP and the mean-field-like (1,2)-TSP, which has its origin
        directly in the mapping of the Hamiltonian circuit problem on the TSP.
        Our data supports previous conjectures that the Euclidean TSP does not
        show signatures of replica symmetry breaking neither in two nor in
        higher dimension. On the other hand the (1,2)-TSP exhibits
        some signature which does not exclude broken replica symmetry, making it a candidate for
        further studies in the future.
    \end{abstract}


    \maketitle

    \section{Introduction}
        The concept of \emph{replica symmetry breaking} (RSB) was introduced
        in the context of spin glasses~\cite{parisi1979infinite,parisi1983order},
        where it has a long history of debate to which models it applies~\cite{Stein2004spinglgasses}.
        RSB is an assumption about the structure of the phase space
        (or ``energy landscape''), which
        leads to the correct results for the Sherrington-Kirkpatrick (SK)
        spin glass~\cite{Talagrand2006parisi}.
        RSB basically means that the phase space is hierarchically structured
        such that two configurations of very similar energy may be far away
        from each other in the configuration space. The phase
        space becomes \emph{complex}.

        The physics-inspired analysis of the phase-space structure has also
        been applied to combinatorial optimization problems, namely problems
        belonging to the class of nondeterministic polymonial
        (NP)-hard~\cite{cook1971complexity,karp1972reducibility,mertens2002computational}
        problems (or the corresponding decision problems belonging
        to the class of NP-complete problems). For NP-hard problems currently
        only algorithms are known which exhibit a worst-case running time which
        grows exponentially with system size. Examples of NP-hard problems are
        satisfiability~\cite{Montanari2008Clusters} and vertex
        cover~\cite{weigt2001minimal}. Here, ensembles are known where replica
        symmetry (RS) breaks at some value of a control parameter~\cite{hartmann2006phase,moore2011,mezard2009}.
        This appears not to be surprising to many researchers because
        intuitively a hard optimization problem  may correspond to a
        non-trivial energy landscape. This prompted many attempts to
        distinguish easy from hard instances or explore the energy landscape of
        such problems~\cite{kirkpatrick1985configuration,cheeseman1991really,gent1996tsp,hartmann2003statistical,smith2010understanding,dewenter2012phase,schawe2016phase}.

        One of the best-known NP-hard combinatorial optimization problems is the
        Traveling Salesperson Problem (TSP)~\cite{etspNP}.
        Somewhat surprisingly, in contrast to the aforementioned problems,
        only indications for RS have been found within studies of some TSP ensembles so
        far~\cite{mezard1986replica,mezard1986mean,sourlas1986statistical,krauth1989cavity}.
        Nevertheless, for these analytical and numerical studies various approximations had to
        be used, somehow questioning the previous claims for RS.

        In this work, by performing computer simulations \cite{practicalGuide} via
        calculating numerically exact ground states \cite{opt-phys2001} and excitations,
        we confirm the previous results for these specific ensembles.
        But on the other hand we show that there is at least one ensemble also for
        the TSP where RSB can not be excluded, namely the $(1,2)$-TSP ensemble~\cite{papadimitriou1993traveling}.
        In particular, in contrast to previous numerical studies, which used heuristics to
        generate tours near the optimum~\cite{sourlas1986statistical,krauth1989cavity},
        we use an exact algorithm to find the true optimum and very specific
        excitations. This approach is facilitated by the combination of
        flexibility and high performance (compared to other exact algorithms for
        the TSP) of linear programming (LP) with branch and cut.
        Combined with the general increase in computing power and the improvement
        of algorithms for TSP optimization, it enables us to simulate comparatively
        large instances.

    \section{Models}
        The Traveling Salesperson problem~\cite{Menger,cook2012pursuit} is defined on a complete
        weighted graph, where the vertices are usually called \emph{cities} and
        the symmetric edge weights $c_{ij} = c_{ji}$ \emph{distances} or \emph{costs}.
        On this graph one searches for the shortest cyclic path through all $N$
        cities, which is called \emph{tour} and can be represented by a set
        of edges $T$. An equivalent representation is through a symmetric
        adjacency matrix $\{x_{ij}\}$ where $x_{ij}=1$ if city $i$ is followed
        by city $j$ on the tour and $x_{ij}=0$ otherwise. The \emph{length} of the
        tour, which we will also call \emph{energy}, is thus
        \begin{align*}
            L = \sum_{\{i,j\}\in T} c_{ij} = \sum_{i} \sum_{j<i} c_{ij} x_{ij}.
        \end{align*}
        Note that an instance of the problem is completely encoded in the
        distance matrix $c_{ij}$.

        To compare two tours $T_1$ and $T_2$, their \emph{distance} or
        \emph{difference} $d$ is defined as the number of edges, which are in
        $T_1$ but not in $T_2$~\cite{kirkpatrick1985configuration}
        \begin{align*}
            d = \sum_{\{i,j\} \in T_1} 1 - x^{(2)}_{ij},
        \end{align*}
        where $x^{(2)}_{ij}$ is the adjacency matrix corresponding to $T_2$.
        Like the link overlap for spin glasses is robust against the flipping
        of \emph{compact} clusters with a low domain-wall energy, this
        observable is robust against partial reversals of the tour. If one
        considered instead the order of the cities in the tour, roughly
        analogous to the spin overlap used for spin glasses, this could
        introduce a difference in the order of $N$ by just changing two links.

        Here, we study four  TSP ensembles to evaluate the influence of the quenched randomness
        on the complexity of the solutions.

        (a) First, the most intuitive and probably the most scrutinized~\cite{beardwood1959shortest,gent1996tsp,percus1996finite,arora1998polynomial,etspNP,Garey1976npcomplete}
        ensemble is the \emph{Euclidean TSP} (ETSP). Here a Poisson point process in a square
        determines the locations of the cities and the distance matrix is
        filled with their Euclidean distances. We use periodic boundary
        conditions. An example for an optimal tour in such a configuration is
        shown in Fig.~\ref{fig:ex:eucOpt}. It is straight forward to generalize
        this in higher dimensions using a Poisson point process in a hypercube
        and the corresponding Euclidean distances.

        (b) The \emph{random link model} (RLTSP)~\cite{kirkpatrick1985configuration,cerf1997random}
        is an approximation of the ETSP, which disregards any correlations of the
        entries in the distance matrix $c_{ij}$ and therefore does not obey,
        e.g., the triangle inequality. For this approximation
        in the statistical physics literature solutions were obtained
        under the premise that replica symmetry holds based on the replica
        method~\cite{mezard1986replica} and cavity method~\cite{krauth1989cavity,cerf1997random,percus1999stochastic}.
        In this work we study the ETSP and RLTSP ensembles in which
        the density of the cities is constant, such that the average optimal
        tour length $L^o \sim N$~\cite{beardwood1959shortest}, i.e., the energy
        is extensive.

        (c) The \emph{$(1,2)$-TSP} is the result of the classical mapping of the Hamilton
        circuit problem (HCP) onto the TSP~\cite{karp1972reducibility}. The HCP
        is whether a cycle visiting every vertex exactly once exists on a given
        graph $G$.
        The mapping from HCP to TSP is simply assigning the distance matrix as
        \begin{align*}
            c_{ij} =
            \begin{cases}
                1, &\text{if $i$ and $j$ are adjacent in $G$,}\\
                2, &\text{otherwise.}
            \end{cases}
        \end{align*}
        A Hamiltonian cycle exists, iff the length of the optimal tour is equal $N$.
        For simplicity sake, the ensemble we are looking at, is derived from
        an Erd\H{o}s-R\'{e}nyi graph (ER) \cite{erdoes1960} where edges occur with
        probability $p=1/N$, which results in an average degree of $1$. Note that
        both limiting cases $p=0$ and $p=1$ are trivial since every tour will
        be optimal with length $2N$, respectively $N$. $p=1/N$ was chosen
        since it is the percolation threshold for the underlying graph ensemble,
        i.e., $G$ exhibits a forest-like
        structure and to form a cycle in the corresponding
        TSP realization almost surely edges non existing in $G$, i.e.,
        distance $2$ in the TSP, need to be used.

        (d) An additional ensemble that we look at is an Euclidean TSP, where the
        cities are arranged on a square lattice with lattice constant $1$
        (STSP). Every city is displaced by at most $1/N$
        in a random direction to avoid degeneracy. An optimal tour in such a
        configuration is shown in Fig.~\ref{fig:ex:sOpt}.
        While this ensemble may appear arbitrary and trivial at first, because it is very
        similar to a grid which is easy to solve, it is actually rather
        nicely motivated. First, the constructions to map the \emph{exact cover} problem onto the
        ETSP~\cite{etspNP,Garey1976npcomplete} result in instances where most
        cities lie on the sites of a square lattice, though not every site is
        occupied. This mapping is the usual way to show that even the ETSP is
        NP-hard.
        Second, historically the ``ts225'' instance of the TSPLIB~\cite{reinelt1991tsplib}
        with 225 cities was solved only 1994 -- three years after the record of
        the largest optimally solved non-trivial instance
        was set to 2392 cities and ten years after its inclusion in the TSPLIB~\cite{applegate1998solution}.
        The empirically hard ts225 instance consists of cities on square
        lattice sites and equidistant cities on straight lines between nearest
        neighbor sites.
        Since we want to look at an easy to define ensemble, we propose the
        slightly disturbed square lattice, which we suspect could show
        typical properties of these square-lattice-like configurations. It turns
        out that open boundary conditions lead to strong finite-size effects,
        since overlaps between two arbitrary tours are coerced at the boundary.
        Therefore, like for the ETSP, we use periodic boundary conditions
        for the STSP ensemble.

        The STSP is obviously a very specific subset of the ETSP. The
        justification to expect a different behavior is that the typical ETSP
        instance might be diluted by entropically favored instances with trivial
        solution space structure, but the solution space structure of the STSP
        subset might look complex.
        Subspaces in the problem domain which behave dramatically
        different are quite common. For example a subspace of the spin glass
        configuration space are ferromagnets, which have a trivial solution space
        structure, while general spin glasses -- at least in high dimensions -- have
        complex ones.

    \section{Methods}
        Like other studies on the solution space structure of different
        optimization problems, we look at excitations~\cite{palassini2000nature,Zumsande2009low,zumsande2008first}.
        To test whether RSB is a possibility, we test a necessary criterion
        introduced in the context of TSP by M\'{e}zard and Parisi in Ref.~\cite{mezard1986mean}.
        A configuration is called \emph{quasi-optimal} if the relative difference
        of its energy $L^*$ to the optimal energy $L^o$ behaves as
        \begin{align}
            \label{eq:crit:relEnergy}
            \frac{L^* - L^o}{L^o} = \mathcal{O}\brac{\frac{1}{N}}.
        \end{align}
        According to Ref.~\cite{mezard1986mean}, in order for replica symmetry
        to be broken, it is necessary that there exist quasi-optimal
        configurations, whose differences to the optimum behaves as
        \begin{align}
            \label{eq:crit:diff}
            d(T^o, T^*) = \mathcal{O}(N).
        \end{align}
        This does not say anything about other configurations which will have
        other distances to the optimum, there will be always a distribution
        of distances to the optimum. This distribution depends on the instances and on the
        system size, similar to the distribution of overlaps in spin glasses \cite{parisi1983order}.
        Thus, for the present analysis, it is not relevant
        whether this distribution of distances converges, or whether the mean converges or whether
        in case of convergence they are self-averaging.
        Intuitively Eq.~\eqref{eq:crit:diff} means, that a finite, i.e., $\mathcal{O}(1)$, energy
        is sufficient to find \emph{some} change of a finite fraction, i.e., $\mathcal{O}(N)$, of
        the system~\cite{palassini2000nature}.
        If this criterion is not fulfilled, we will conclude that RS holds.

        Furthermore, we have to ensure that some kind of order exists in the
        ground state. Consider for example a system, where every edge has equal
        length.  The solution space structure is like a paramagnet, i.e.,
        trivial since every tour is of identical length. On the other hand,
        this system also fulfills the criterion Eq.~\eqref{eq:crit:diff}. While
        a random tour and the optimal tour in this degenerate ensemble behave
        the same in every aspect, this is not true for the $(1,2)$-TSP, where a
        random tour has $\mathcal{O}(1)$ edges of length \emph{one} but an
        optimal tour has $\mathcal{O}(N)$ edges of length \emph{one}.
        This distinguishes order from disorder. In more detail, our
        measurements show the actual number of length \emph{one} edges for the
        $(1,2)$-TSP is $0.4240(8) N$, thus, corresponding to an ordered ground
        state. We obtained this constant by using the
        Beardwood-Halton-Hammersley constant $\beta$, which will
        be scrutinized in the beginning of Sec.~\ref{sec:results}. For the
        $(1,2)$-TSP $\beta$ is the mean length of the edges
        constituting the tour and since all edges in the $(1,2)$-TSP ensemble
        are of length $1$ or $2$, $\beta-1$ is the fraction of length $2$ edges
        in the optimal tour.
        The ETSP shows a very similar behavior~\cite{vannimenus1984statistical}.
        We will further show that the STSP on the other hand, while fulfilling the criterion
        Eq.~\eqref{eq:crit:diff}, behaves still trivial and the fulfillment of
        the criterion is caused by a high degeneracy.

        Note that degeneracy alone does not mean that a solution space
        structure is trivial, since the degenerate solutions may be contained
        in one big cluster, at least in the thermodynamic limit.
        Famous examples, where this is the case include the two-dimensional
        Ising spin glass with $\pm 1$ couplings~\cite{hed2001spin} and the
        satisfiability problem in the range of few
        constraints~\cite{Monasson1997statistical}.

        Anyway, for the cases where we can not rule out RSB, we can not
        reach a definitive conclusion since RSB is a more complex phenomenon not only
        caught by one quantity of interest.  However, we can identify cases which
        might be worthwhile to study in more detail to determine whether they
        are RS or RSB, or exhibit a complex behavior in another way.

        Going from measurable quantities to algorithms, to solve numerically
        any instance of the TSP, the following integer program,
        i.e., an LP with additional integer constraints Eq.~\eqref{eq:int},
        can be used~\cite{dantzig1954solution}

        \begin{align}
            \label{eq:min}
            &\text{minimize}     & \sum_i \sum_{j<i} c_{ij} x_{ij} \\
            \label{eq:inout}
            &\text{subject to}   & \sum_{j} x_{ij}                       &= 2&            & i = 1,2,...,N \\
            \label{eq:sec}
            &                    & \sum_{i \in S, j \notin S} x_{ij}     &\ge 2&          &\forall S \subset V,\\
            \label{eq:int}
            &                    & x_{ij}                                &\in \{0,1\}
        \end{align}
        where $x_{ij}$ is the searched for adjacency matrix defining the tour,
        $V$ is the set of all cities and $S$ a proper, non-empty subset of $V$.
        Eq.~\eqref{eq:min} minimizes the tour length, Eq.~\eqref{eq:inout} ensures
        that the number of incident edges into every city is two, such that
        the salesperson enters every city once and leaves it again. Eq.~\eqref{eq:sec}
        are the \emph{subtour elimination constraints} (SEC), which prevent the tour
        to fragment into multiple not-connected subtours.

        As a technical detail, we use fixed point data types for the distances.
        This discretization means effectively that the entries of the cost
        matrix are rounded and can therefore lead to different results than
        exact Euclidean distances, however this is a fundamental problem of
        any computer simulations. Tests with different precisions did not show
        any systematic and notable influence on the mean values, such that we are confident that
        no systematic error is introduced by this choice.
        We use Concorde~\cite{applegate2003implementing} to generate optimal tours,
        which implements the LP from Eq.~\eqref{eq:min} to~\eqref{eq:int} at
        its core but also extends it with additional constraints and heuristics
        to speed up the solution process.

        Note that optima found with this method are not necessarily drawn
        uniformly from all existing optima and we do not perform unbiased
        ground-state sampling. However, most of our ensembles are not degenerate anyway. And
        in the case of the $(1,2)$-TSP, the only
        ensemble with many optima studied by us, we tested whether this possible
        bias has influence on our results. Therefore we applied random perturbations
        on the edge lengths to lift the degeneracy, which was tested on other
        models to result in uniform, unbiased sampling of the optima \cite{Amuroso2004domain,Ohr2018exact}.
        This procedure yielded within errorbars the same results as the
        degenerate ensemble, such that we are confident that any possible bias
        of the optimum selection does not have considerable influence on our
        results.

        To construct the excitations $T^*$, we modify the linear program
        formulation using the obtained optimal tour $T^o$.
        This allows us to construct excitations with very specific properties.
        Since we want to check the criterion Eqs.~\eqref{eq:crit:relEnergy} and
        \eqref{eq:crit:diff}, we construct a very specific integer
        program which fixes Eq.~\eqref{eq:crit:relEnergy} to be
        fulfilled and maximizes Eq.~\eqref{eq:crit:diff}. If the
        problem is RS, the result should show the criterion to be violated.

\begin{figure}[bhtp]
            \centering
            \subfigure[\label{fig:ex:eucOpt} ETSP, optimal]{
                \includegraphics[width=0.31\linewidth]{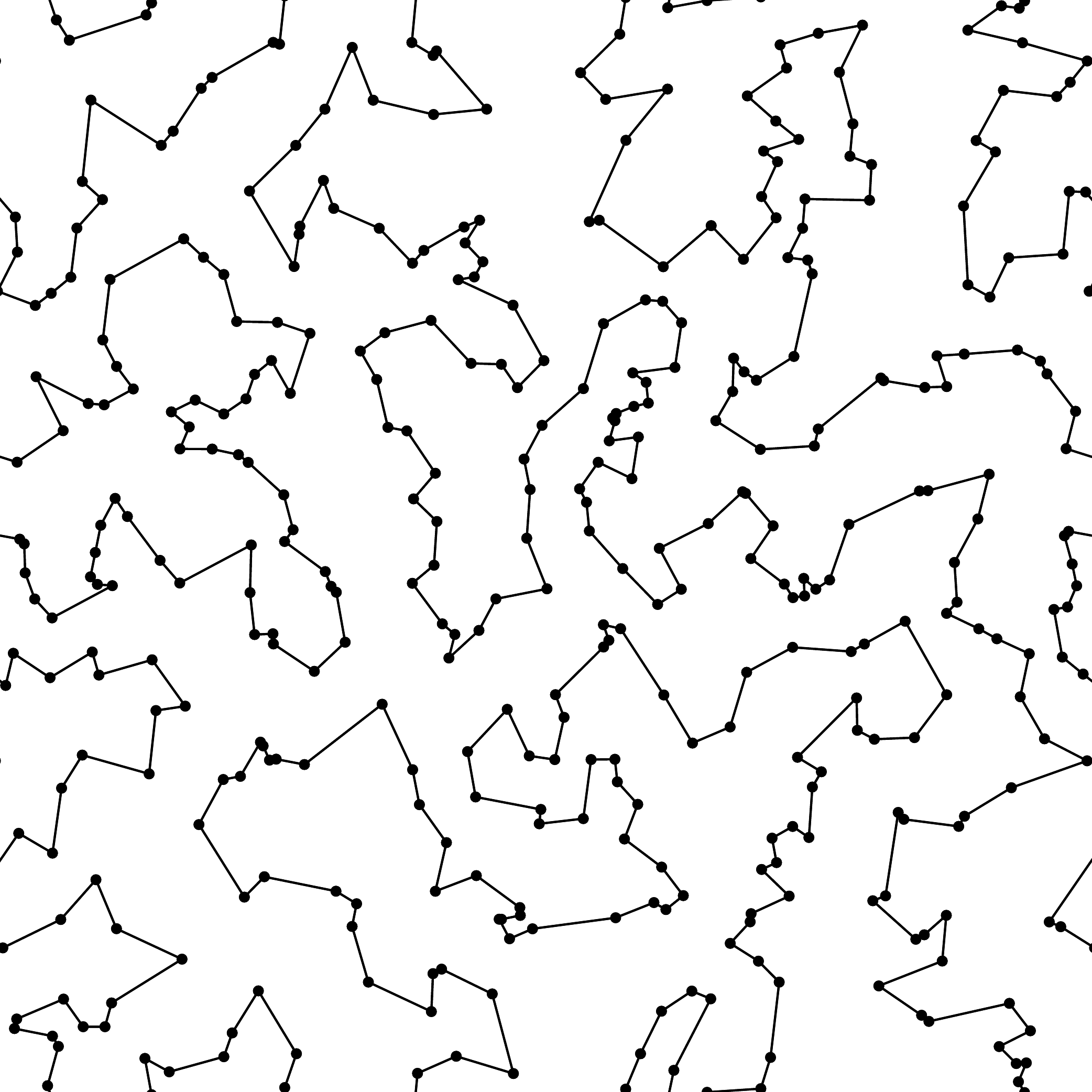}%
            }
            \subfigure[\label{fig:ex:eucMaxDiff} ETSP, MaxDiff]{
                \includegraphics[width=0.31\linewidth]{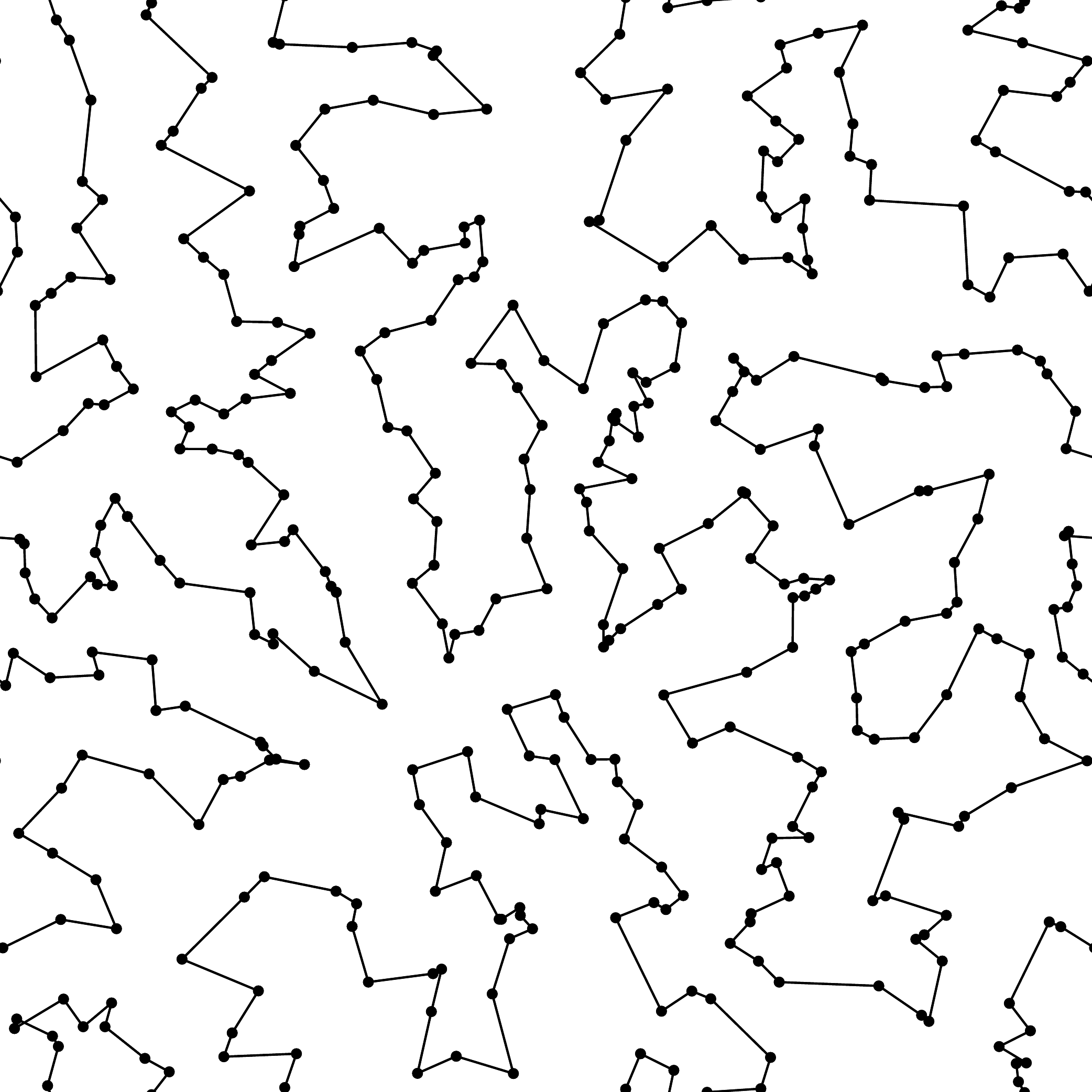}%
            }
            \subfigure[\label{fig:ex:eucDiff} ETSP, difference]{
                \includegraphics[width=0.31\linewidth]{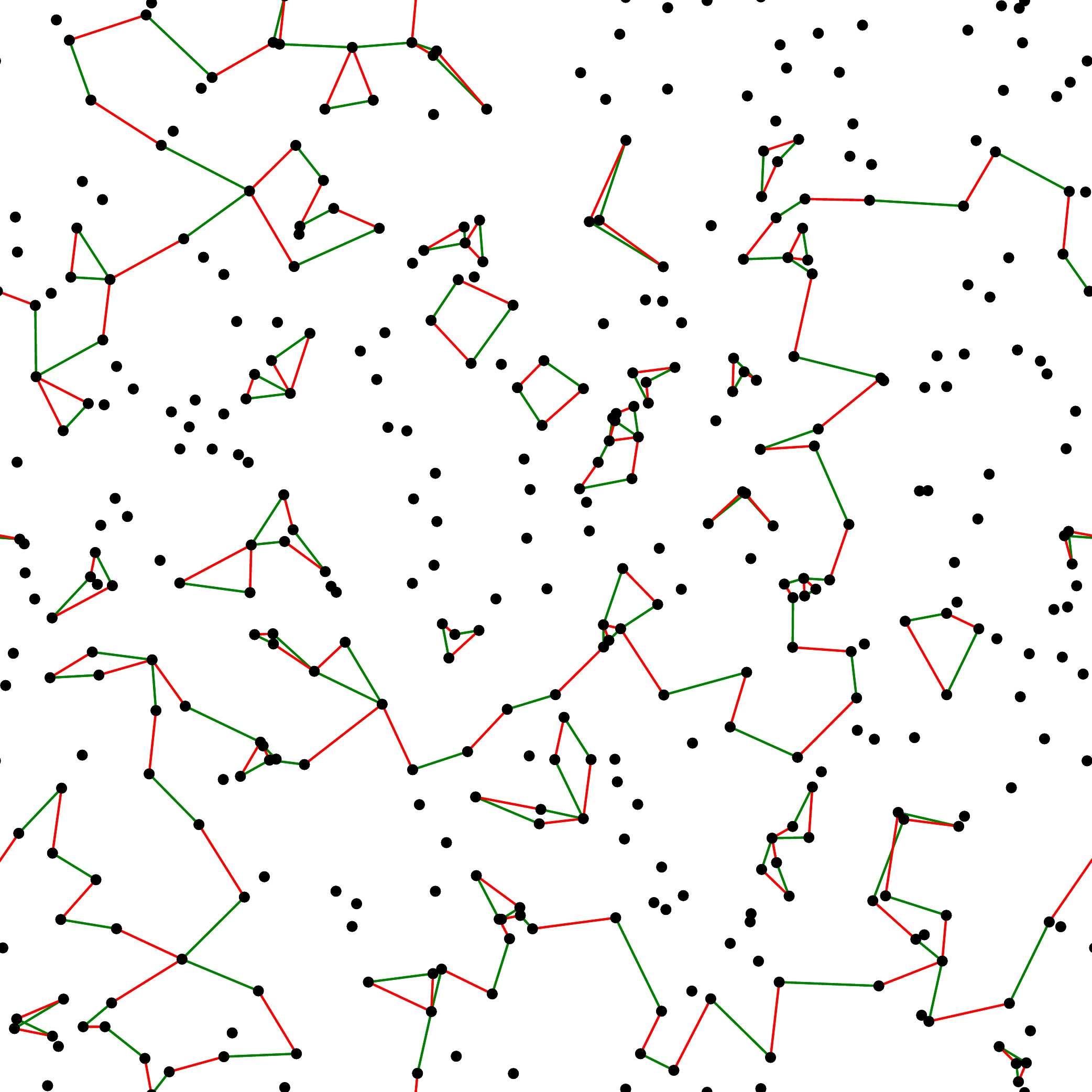}%
            }

            \subfigure[\label{fig:ex:sOpt} STSP, optimal]{
                \includegraphics[width=0.31\linewidth]{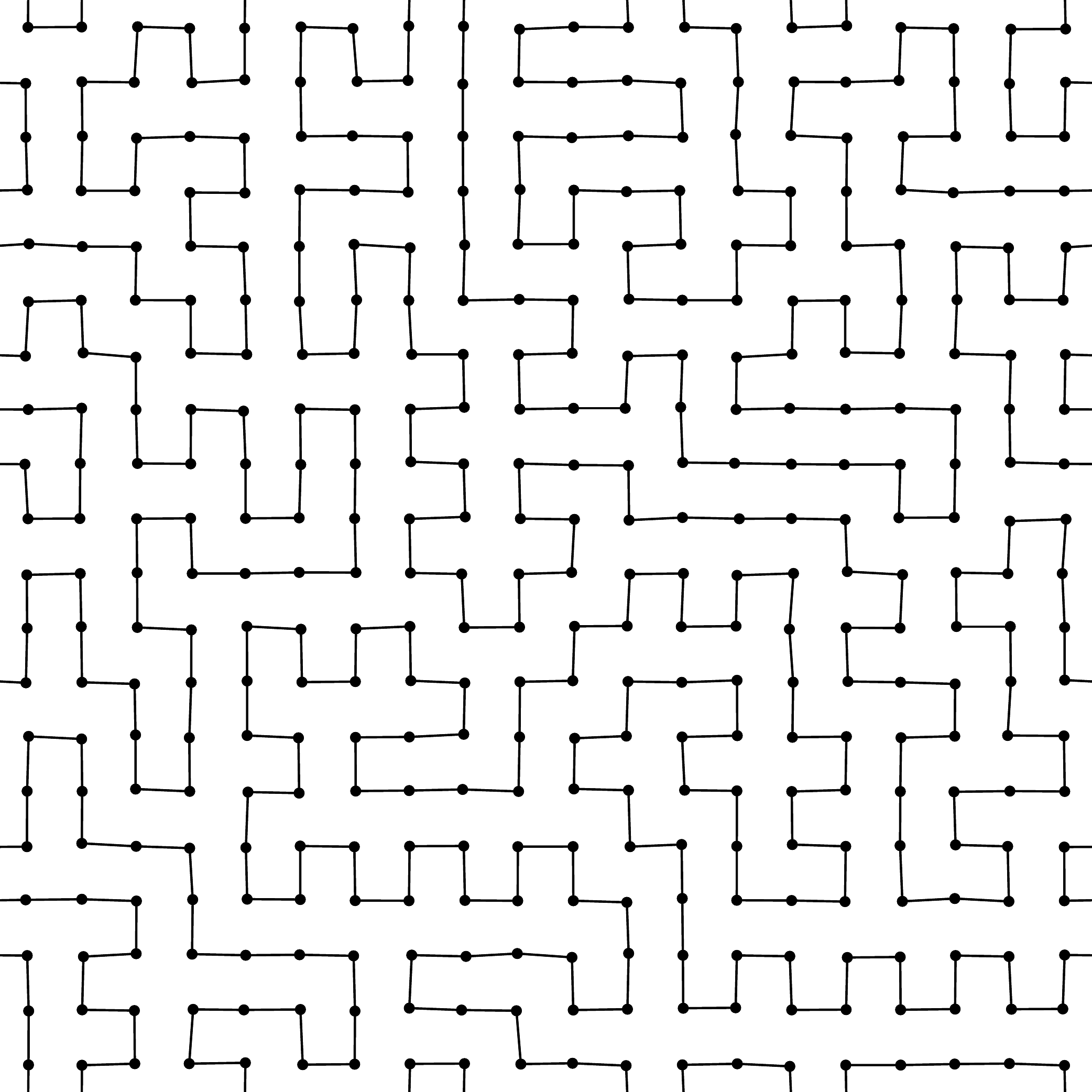}%
            }
            \subfigure[\label{fig:ex:sMaxDiff} STSP, MaxDiff]{
                \includegraphics[width=0.31\linewidth]{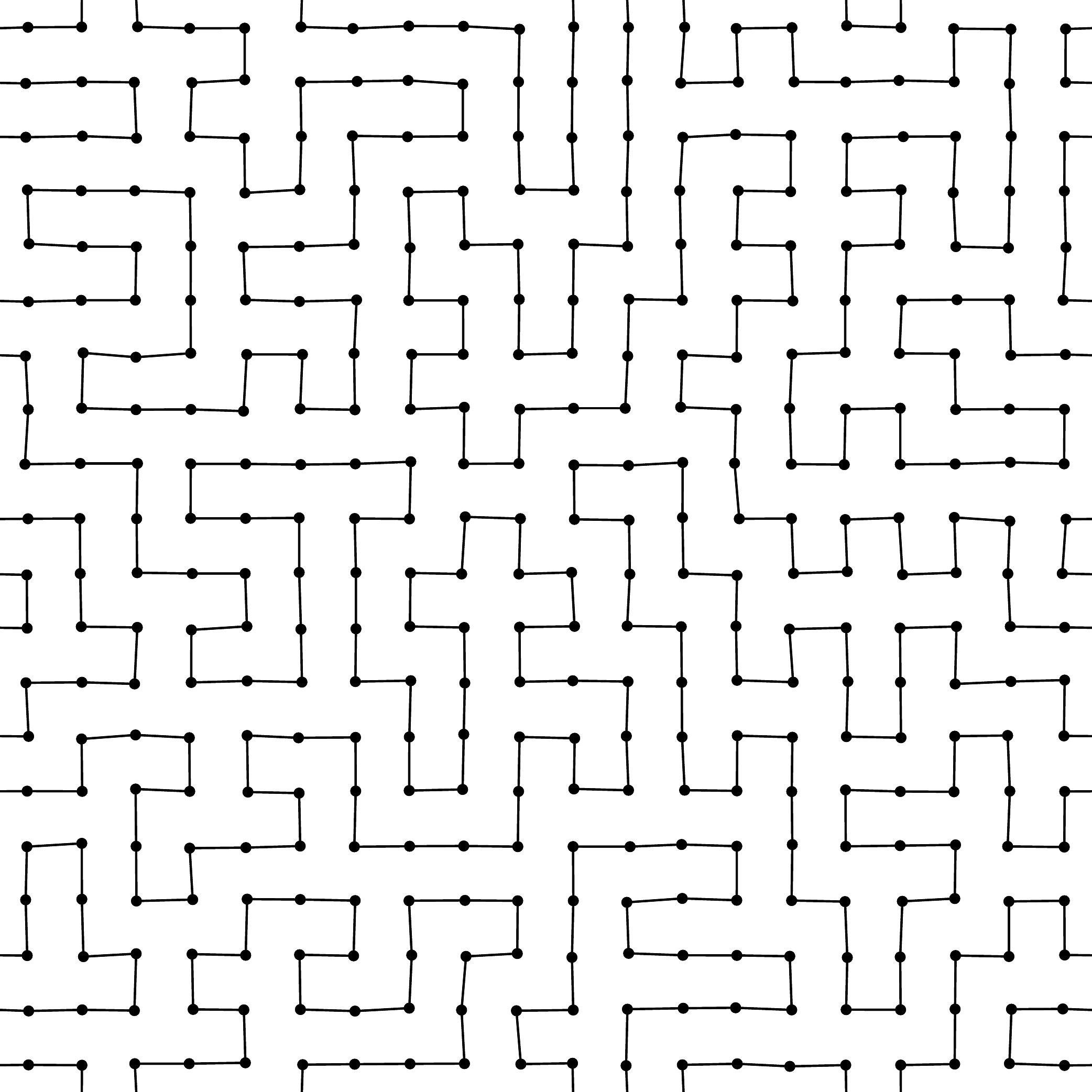}%
            }
            \subfigure[\label{fig:ex:sDiff} STSP, difference]{
                \includegraphics[width=0.31\linewidth]{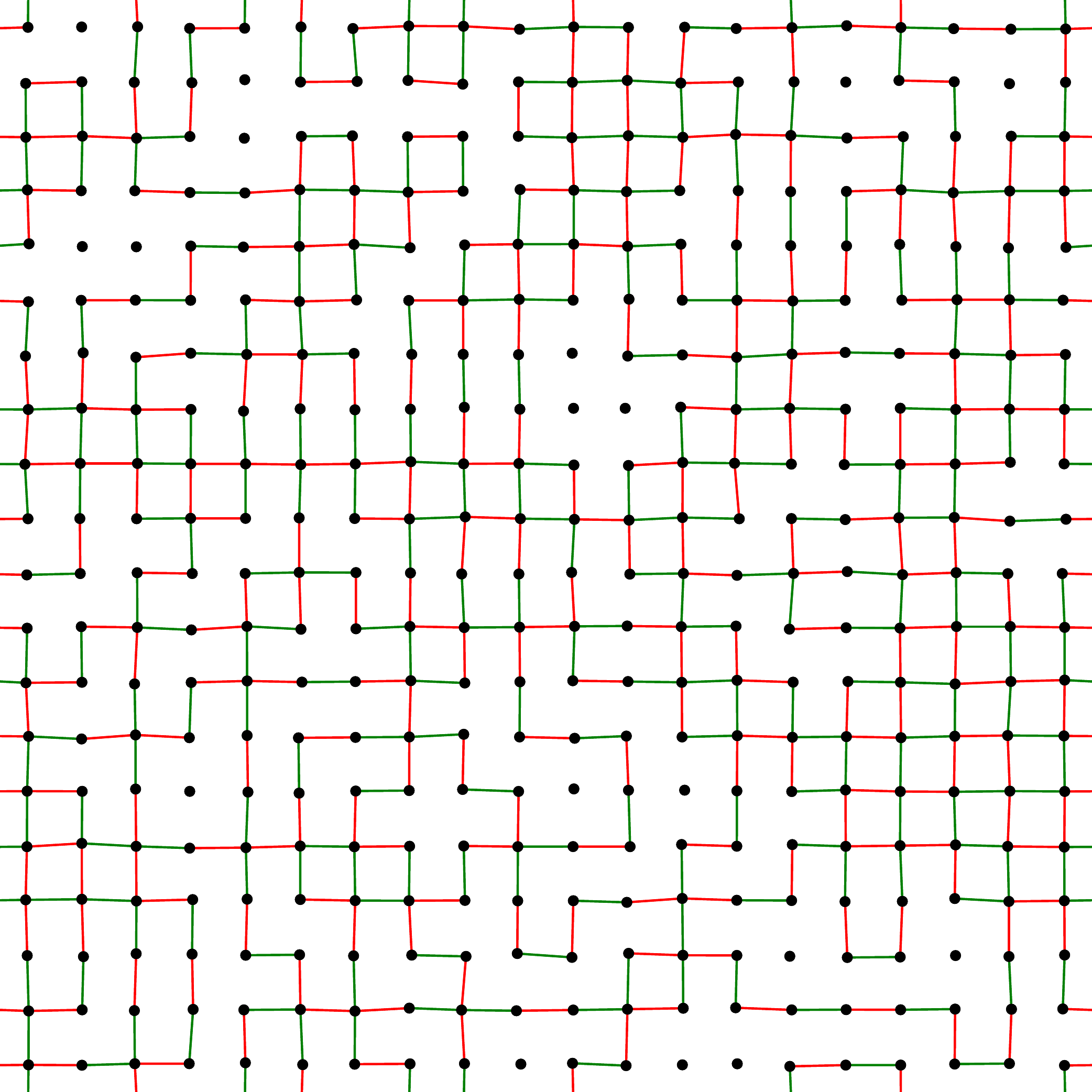}%
            }

            \caption{\label{fig:ex} (color online)
                \subref{fig:ex:eucOpt} and \subref{fig:ex:eucMaxDiff}
                show a configuration with $N=400$ of the ETSP.
                Left is the optimal tour, right the MaxDiff excitation with
                $d=129$ difference to the optimum.
                \subref{fig:ex:sOpt} and \subref{fig:ex:sMaxDiff} show the
                optimal and excited tour for a STSP realization.
                \subref{fig:ex:eucDiff} and \subref{fig:ex:sDiff} show the
                difference between the optimum and the excitation of the
                respective instances, red edges are removed, green are added
                for the excitation.
            }
        \end{figure}

        So we fix the allowed energy difference $L^* - L^o = \epsilon$
        to a constant, which will lead to the desired relative energy difference
        Eq.~\eqref{eq:crit:relEnergy} if the energy is extensive.
        For this reason our definitions of the ensembles are formulated in a way
        that leads to extensive energy, i.e., $\avg{L^o} \sim N$.
        Within this excitation energy window $\epsilon$, the number of common edges with
        the optimal tour $T^o$ needs to be minimized to maximize the distance
        of the configurations. Thus replacing the objective with
        \begin{align}
            \label{eq:maxDiff}
            \text{minimize}     & \sum_{\{i, j\} \in T^o} x_{ij}
        \end{align}
        and adding the additional constraint
        \begin{align}
            \label{eq:eps}
            \sum_i \sum_{j<i} c_{ij} x_{ij}       &\le L^o + \epsilon
        \end{align}
        results in a suitable LP. We will call this LP \emph{MaxDiff}.
        Technically, we used a custom implementation of the LP. We
        used CPLEX~\cite{cplex} as the LP solver and for branch and cut.
        Two exemplary solutions of this LP are visualized in Fig.~\ref{fig:ex}
        in comparison to the optimal tours.

    \section{Results}
    \label{sec:results}
        We performed the calculation of optimum and excited tours for the four ensembles
        ETSP, RLTSP, (1,2)-TSP, and STSP, for various system sizes ranging from $N=64$
        to $N=1448$ cities.  All results are averaged over few 100 realizations of the disorder.

        Due to the hardness nature of the TSP and our exact solution approach, some
        realizations of the largest system sizes take far more computational resources than most
        realization and could
        not be solved in reasonable time respectively memory.
        If we just omitted the unsolved instances from our results, the results
        would be subject to selection bias since the hardest realizations (for
        the used algorithm) are excluded from the means, which can lead to
        systematical errors -- especially since we are interested in properties
        linked to hardness. To ensure that the results are not tainted by such
        systematic errors, we perform a very conservative error estimation.
        The basic idea is that we determine intervals of possible values for
        the means, where unsolved instances enter with their minimum and
        maximum possible values, thus taking care of these systematic errors.
        As we will see in the results section, these intervals are very small, showing that the
        unsolved instances have no significant effect, which is visible also when comparing
        fits which used lower and upper possible means.
        Nevertheless, in detail, there are two ways in which the optimization might fail,
        which we treated differently.
        In the first case, already the optimal tour, i.e., the groundstate, can
        not be found for a given realization. System sizes for which this happened
        for at least one realization are omitted completely from our analysis.
        In other words, our data contains only system sizes where we always
        found the optimum tour for all instances of this system size and ensemble.
        In the second failure case, we found the optimal tour Eq.~\eqref{eq:min}
        for an instance, but were not able to determine the excitation
        Eq.~\eqref{eq:maxDiff}. In this case we can often use intermediate
        results of the branch-and-cut procedure to estimate upper and lower
        bounds. The upper bound of $d$ is always available as the solution
        of the LP relaxation, i.e., the solution of the LP defined by
        Eqs.~\eqref{eq:maxDiff}, \eqref{eq:inout}, \eqref{eq:sec} and \eqref{eq:eps}
        without the integer constraints. In fact, the bound is even tighter,
        as we can round down the relaxation solution to the next integer.
        The lower bound of $d$ is available if the branching procedure produces
        an integer solution\footnote{
            To find an integer solution during the branching, a strategy with an
            emphasis on the integrality constraints instead of optimality
            should be used. Integer programming libraries often allow to choose
            such a strategy, e.g., in CPLEX with the \emph{MIPEmphasis} parameter
            or in Gurobi with the \emph{MIPFocus} parameter. We also observed that
            this focus does lead typically not only to non-optimal integer
            solutions, which can be used as bounds, but also to faster termination
            of the algorithm with an optimal integer solution.
        },
        otherwise it is assumed as the minimum possible value, i.e., $d=0$.
        Similarly, we can estimate bounds on the relative energy difference
        as it is bounded by $0$ and $\varepsilon/L^o$.

        The range of sizes $N$ we study for each TSP ensemble is determined by the largest size
        for which we could calculate always the optimum and obtain results with sensibly small
        variations due to the inclusion of instances where the excitation could not be obtained.
        We used as a criterion than not more than one instance of the second failure kind occurs
        which comes \emph{without} an estimate for the lower bound.
        On the other hand, failures of the second kind, coming \emph{with} upper and lower bounds are quite
        benign, since the bounds are typically reasonably tight, such that
        data points are shown, for which, for the largest sizes, up to $10\%$ of the $100$ (or $200$
        depending on the ensemble) samples belong to this failure category.
        All fits performed in the remainder of this study are done twice. Once
        using the upper bounds and once using the lower bounds. The results are
        always compatible within the statistical errors. All fit results shown
        in the following are obtained from using the upper bound, since it is
        usually tighter than the lower bound.

        \begin{figure}[bhtp]
            \centering
            \includegraphics[scale=1]{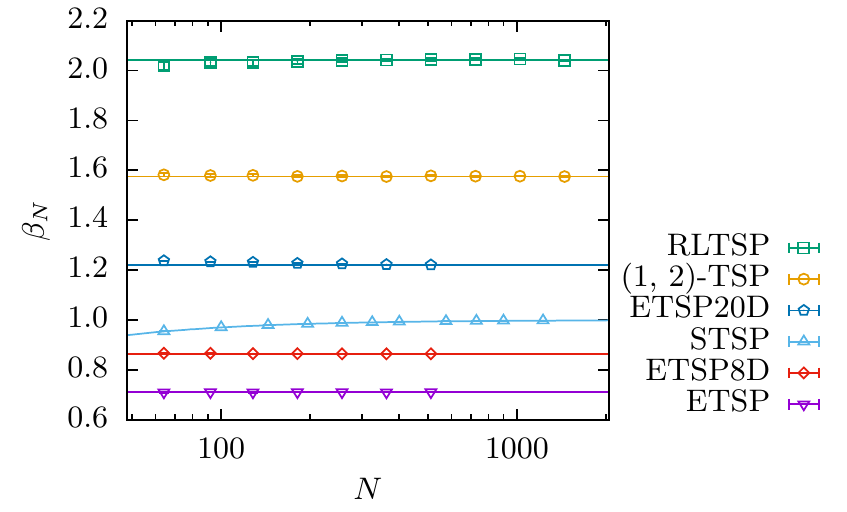}%
            \caption{\label{fig:beta}
                Finite-size Beardwood-Halton-Hammersley constant $\beta_N$ measured at different
                system sizes $N$. Except for the STSP, all lines show the mean
                value calculated from the results for $N \ge 200$. For STSP the line is a fit to
                $\beta_N = \beta + aN^{b}$ to extrapolate the asymptotic $\beta$,
                which yields $a=-2.62(5)$ and $b=-0.971(4)$.
                The extrapolated values for $\beta$ are tabulated in table \ref{tab:beta}.
            }
        \end{figure}

        We start the presentation of the results with the behavior of the optimum tour length.
        Here only data for system sizes $N$ is included where an optimum was found for all instances of
        this size $N$.
        For the ETSP it is well known that the
        mean optimal length $\avg{L^o}$ through $N$ cities placed on
        a unit square by a Poisson point process approaches a limit value for
        large $N$, if scaled appropriately
        \begin{align}
            \label{eq:beta}
            \lim_{N\to\infty} \avg{L^o}/\sqrt{N} = \beta.
        \end{align}
        This constant $\beta$ is the Beardwood-Halton-Hammersley constant \cite{beardwood1959shortest}
        and some estimates for its value exist \cite{percus1996finite,Jacobsen2004traveling,applegate2010using}.
        Similarly, such a constant should
        exist for the random link model. For the pseudo one-dimensional case, it is
        even known exactly \cite{Wastlund2010mean}. For the STSP and (1,2)-TSP
        the authors are not aware of previous work, but it is easy to recognize
        that the optimal tour in the STSP traverses $N$ horizontal or
        vertical edges, which should have each a length of $1$ for large $N$. Thus,
        we expect the corresponding constant to be $\lim_{N\to\infty} \avg{L^o}/N = 1$.

        Comparing these expectations to our data serves as a crosscheck to
        establish some level of confidence in our data. For the (1,2)-TSP case
        these are novel results.
        In Fig.~\ref{fig:beta} the rescaled mean optimal tour lengths $\beta_N=\avg{L^o}/\sqrt{N}$ are
        plotted. Generally the finite-size effects are small, such
        that we determine estimates for $\beta$ simply as the average of all datapoints
        $N \ge 200$. Except for the STSP this works reasonably well. Since the
        STSP shows the largest finite-size effect, we use an offsetted power law
        $\beta_N = \beta + aN^{b}$ to extrapolate the measurements. The results
        of this analysis are shown in Table~\ref{tab:beta} together with
        the currently best known values for this constant. Since we do not have
        a good model to extrapolate the values, the given errorbars are only
        statistical
        and do not account for errors in the extrapolation.
        Considering this, our estimates are reasonably close to the expectations.

        \begin{table}[htb]
            \caption{\label{tab:beta}
                Beardwood-Halton-Hammersley constants $\beta$ for different
                ensembles of the TSP determined from our data and the current
                best estimates for their actual values. Note that we only need
                to find the groundstate and not the excitation for this analysis
                such that we can show larger system sizes than in later
                results.
                The literature values marked by an asterisk * are values
                according to a large $d$ conjecture for the RLTSP, which
                should coincide in this limit with the ETSP \cite{cerf1997random},
                Therefore, a perfect agreement is not expected.
            }
            \begin{ruledtabular}
                \begin{tabular}{lll}
                     & \multicolumn{1}{c}{$\beta$ (measured)} & \multicolumn{1}{c}{$\beta$ (literature)}\\[0.05cm]
                    \hline
                    \noalign{\vskip 0.1cm}
                    ETSP, $d=2$                      & $0.7112(6)$ & $0.712403(7)$ \cite{applegate2010using}\\ 
                    ETSP, $d=8$                      & $0.8645(3)$ & $0.8531*$ \cite{cerf1997random}\\ 
                    ETSP, $d=20$                     & $1.2218(2)$ & $1.2093*$ \cite{cerf1997random}\\ 
                    RLTSP, $d=1$                     & $2.044(3)$  & $2.0415..$ \cite{Wastlund2010mean}\\ 
                    STSP                             & $1.0005(2)$ & $1$\\ 
                    (1, 2)-TSP (ER, $p=\frac{1}{N}$) & $1.5760(8)$ & - \\
                \end{tabular}
            \end{ruledtabular}
        \end{table}

\begin{figure}[bhtp]
            \centering
            \includegraphics[scale=1]{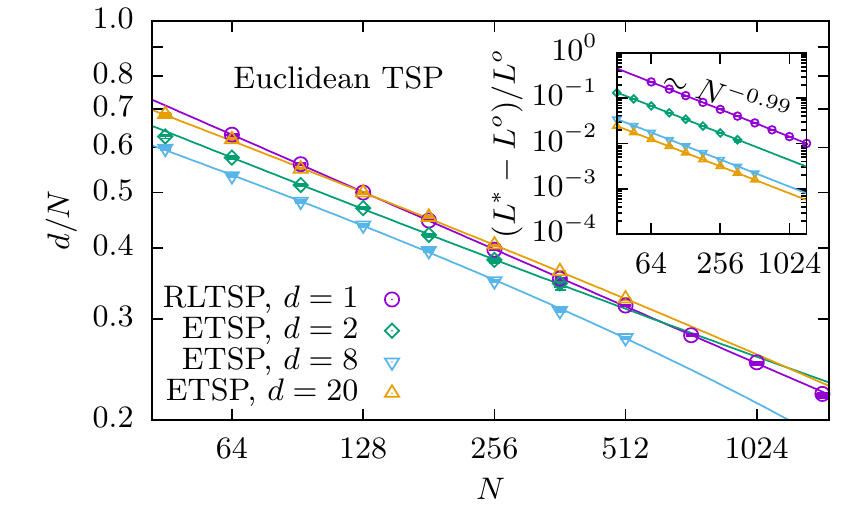}%

            \caption{\label{fig:etsp}
                The relative difference of the optimum and the MaxDiff
                excitation decreases as a power law with the system size $N$. Results for the ETSP for
                various dimensions and the RLTSP, which is believed to show the same behavior, are shown.
                Its exponent dependents on the ensemble. For large $N$ the
                difference $d/N$ vanishes which is a hint for replica symmetry
                and a trivial solution space structure. The inset shows that the
                premise Eq.~\eqref{eq:crit:relEnergy} is fulfilled.
                The higher dimensional cases have a 10 times larger $\epsilon$.
                The bounds of the value are visualized as filled boxes and the
                statistical errors as errorbars, note that both are always
                smaller than the symbols. The smallness (hardly visible) of the
                boxes indicate that the few instances where only bounds for the
                distance $d$ could be obtained have basically no influence as
                the relative difference between upper and lower bound is always
                less than $2\%$. The exponent and offset, which
                is always compatible with 0, are obtained by fits to
                $\frac{d}{N} = a N^{b} + D^\infty$ and shown in
                Table~\ref{tab:measured}.
            }
        \end{figure}

        Next, the results for the MaxDiff excitation simulations for the
        two-dimensional ETSP are shown in Fig.~\ref{fig:etsp}.
        We found a $1/N$ behavior of the relative energy difference (inset) as
        required  by Eq.~\eqref{eq:crit:relEnergy}.
        Nevertheless the difference $d$ of the tours also vanishes in
        the large $N$ limit as a power law, thus the behavior of Eq.~\eqref{eq:crit:diff}
        is not found. Therefore, to change a finite fraction of an
        infinite system, a finite energy $\epsilon$ does not suffice for this ensemble.
        Thus, the results do not show the signature of replica symmetry
        breaking, hinting at a trivial solution space structure. This is
        consistent with previous studies \cite{mezard1986replica,sourlas1986statistical,krauth1989cavity,cerf1997random,percus1999stochastic}
        using the RLTSP as an approximation for the ETSP. They used, e.g., the
        cavity method to estimate some properties and compared them to tours
        obtained by heuristics (for smaller system sizes $N$) leading to the
        claim that the ETSP is replica symmetric. Also our results actually
        for the RLTSP, shown in Fig.~\ref{fig:etsp}, lead to the same
        conclusion and confirm the previous results.

        For spin glasses, the energy landscape becomes complex and exhibits many features
        of RSB in high enough dimensions. Above the upper critical
        dimension the system is believed to
        behave~\cite{harris1976,katzgraber2005ATlinie,katzgraber2009ATline,moore2011ATline}
        like the mean-field SK model~\cite{parisi1979infinite,parisi1983order},
        corresponding to RSB. This motivated us to investigate the ETSP for
        high dimensions as well.  Our results for the $8$-dimensional and $20$-dimensional ETPS
        are also shown in Fig.~\ref{fig:etsp}. Evidently, even for higher dimensions the same behavior
        indicates that RS is present. Thus a
        simple increase in dimensionality does apparently not change the
        behavior regarding replica symmetry much. This is in strong contrast to
        spin glasses. On the other hand, this is not too surprising, because the ETSP allows
        for connections between all cities, i.e., is long-range and in this sense
        mean-field like in all dimensions.

\begin{figure}[bhtp]
            \centering
            \includegraphics[scale=1]{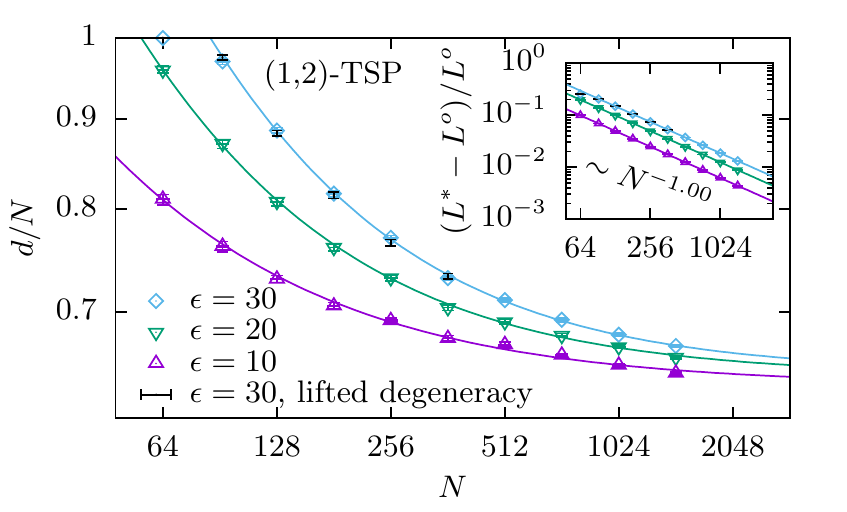}%
            \caption{\label{fig:12}
                Statistics of the $(1,2)$-TSP for a connectivity of $Np = 1$.
                The MaxDiff constraints with the finite excitation energy
                $\epsilon \in \{10, 20, 30\}$ are used for the three curves
                respectively.
                The distance of the excitation to the optimal tour is extrapolated
                with an offsetted power law ansatz $\frac{d}{N} = a N^{b} + D^\infty$.
                The fit parameters are obtained for $N>100$ and are shown in
                Table~\ref{tab:measured}.
                All three result in a convergence to a finite
                $D^\infty$ for large $N$, i.e., a finite fraction. The possibility
                of RSB can therefore not be excluded.
                The inset shows the relative energy difference of the optimum
                and the excitation, showing nearly a perfect $1/N$ form, as
                required by the RSB criterion.
                The bounds of the value are visualized as filled boxes and the
                statistical errors as errorbars, note that both are always
                smaller than the symbols. The best estimate, i.e., the
                upper bound, is used for fits.
            }
        \end{figure}

        Next, we will look at an ensemble which is closer to a direct mapping
        from the Hamilton circuit, which is usually used to prove the TSP
        NP-complete.
        The mapping creates an instance of the $(1,2)$-TSP. For three
        tested values of the finite excitation energy $\epsilon \in \{10, 20, 30\}$,
        we calculated the difference between the optimal and excited tours $d$,
        shown in Fig.~\ref{fig:12}.
        First, see inset, the relative energy
        difference decreases as $1/N$ as required by Eq.~\eqref{eq:crit:relEnergy}.
        The measured difference $d$
        does not follow a pure power law, but seems to converge to a non-zero
        offset. Extrapolating the difference for large $N$ with
        $\frac{d}{N} = a N^{b} + D^\infty$ (cf.~Ref.~\cite{palassini2000nature})
        leads to offsets for each value of $\epsilon$, which are reasonably close to
        the most accurate value we obtained $D^\infty = 0.645(2)$ and exponents
        close to $b=-1$. All values are shown in Table~\ref{tab:measured}.
        Note that for small system sizes $N$ finite-size effects are visible, where $\epsilon$ is of
        the order of the optimal length and the excitation can differ in every
        single edge. Therefore, the difference is clamped at $d/N = 1$. For
        larger system sizes $N$ this does not seem to play a role anymore. To reduce the
        influence of this finite-size effect, the fits for larger $\varepsilon$
        disregard the small system sizes $N<128$ for $\varepsilon = 30$ and
        $N<64$ for $\varepsilon = 20$. In particular,
        different values of $\epsilon$ lead to consistent results.
        According to the criterion Eq.~\eqref{eq:crit:diff} our results
        can not exclude the possibility that replica symmetry is actually
        \emph{broken} for this ensemble.

        To further test these results, we conducted simulations above the percolation
        threshold, for $p=3/N$, and below the threshold for $p=1/2N$. The
        results exhibit qualitatively the same behavior (not shown),
        but with different values of the asymptotic $D^\infty$. Apart from
        the limits $p\to 0$ and $p\to 1$, where every tour is optimal, the
        precise structure of the graph does not seem to have a critical influence
        on this result.

        To exclude that the degeneracy has a special influence on our results,
        we lift the degeneracy by adding a slight perturbation on each
        edge. Therefore we scale the edge weights and $\epsilon$ by $\cdot 10^5$ and
        add a random disturbance $U(-10, 10)$ to each edge. Except for a
        vanishing degeneracy, this procedure also does not change the results
        beyond statistical errors, which are indicated as additional black
        errorbars below a selection of datapoints for $\varepsilon = 30$ in
        Fig.~\ref{fig:12}.

\begin{figure}[bhtp]
            \centering
            \includegraphics[scale=1]{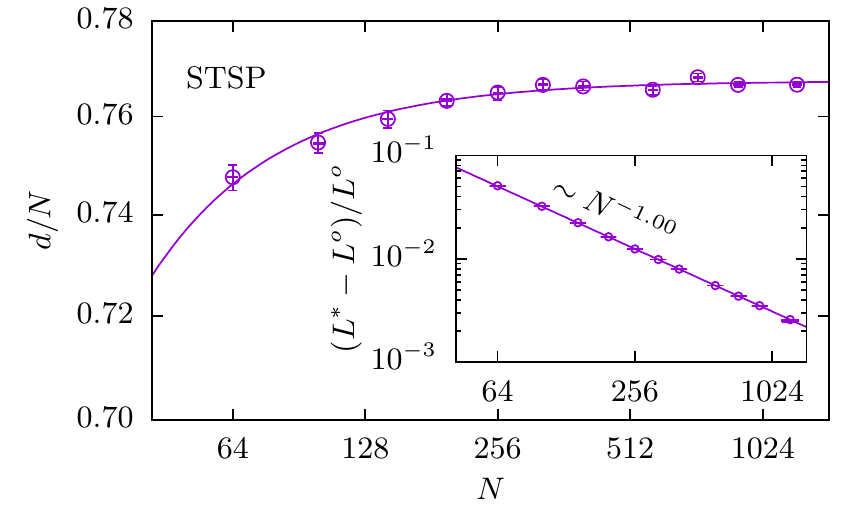}%
            \caption{\label{fig:stsp}
                For the STSP the relative difference $d/N$ converges to a finite
                value, which means that finite energy is sufficient to change a
                macroscopic part of the system. The value it converges to
                is estimated by an offsetted power law ansatz $\frac{d}{N} = a N^{b} + D^\infty$
                and fulfills the criterion Eq.~\eqref{eq:crit:diff}.
                The exponent and offset are shown in Table~\ref{tab:measured}.
                The inset shows that Eq.~\eqref{eq:crit:relEnergy} is fulfilled.
                The bounds of the value are visualized as filled boxes and the
                statistical errors as errorbars, note that the bounds are always
                smaller than the symbols. The best estimate, i.e., the
                upper bound, is used for fits.
            }
        \end{figure}

        The last ensemble we study is the STSP, which is a very special
        subspace of the ETSP configuration space.
        The STSP, where cities are placed on a square lattice and are displaced
        by a distance proportional to $1/N$, does show a qualitatively very
        different behavior to the ETSP. In contrast to the ETSP case, the
        difference does not follow a pure power law, but seems to converge to a
        non-zero offset. But different than the $(1,2)$-TSP case, it approaches
        the limiting value from below.
        In Fig.~\ref{fig:stsp} this behavior is fitted with an offsetted
        power law $\frac{d}{N} = a N^{b} + D^\infty$. While according to
        criterion Eq.~\eqref{eq:crit:diff} this is not compatible with the
        trivial behavior of RS, it is rather easy to see that this
        is an effect caused by high degeneracy. In fact, large realizations
        basically look like a square lattice, where many tours which do not use
        diagonals have almost the same lengths. This is compatioble with the value of
        $\beta$ which is apparently $1$ (cf. Table~\ref{tab:beta}). The
        slight displacements avoid perfect degeneration, but are not strong
        enough to destroy this effect and thus no ordered phase can be observed.
        Thus, the system behaves like a paramagnet, where many solutions are indistiguishable
        close in energy.
        Note that a displacement by a fixed amount, e.g., $5\%$ of the lattice
        constant, does lead to the same trivial behavior as the ETSP before (not shown).
        The same is true for a diluted square lattice, where a fixed fraction
        of sites is removed (also not shown). We therefore conclude that the
        energy landscape is most likely trivial.

        \begin{table}[htb]
            \caption{\label{tab:measured}
                Values of the fit parameters extrapolating the behavior of $d/N$.
                Interestingly all ensembles converging to a finite value of
                $D^\infty$ show an exponent close to $b = -1$.
            }
            \begin{ruledtabular}
                \begin{tabular}{lllc}
                     & \multicolumn{1}{c}{$b$} & \multicolumn{1}{c}{$D^\infty$} & \multicolumn{1}{c}{RS}\\[0.05cm]
                    \hline
                    \noalign{\vskip 0.1cm}
                    ETSP, $d =  2$              & $-0.32(8)$   & $\phantom{-}0.04(10)$ & \checkmark   \\ 
                    ETSP, $d =  8$              & $-0.21(4)$   & $-0.18(9)$            & \checkmark   \\ 
                    ETSP, $d = 20$              & $-0.27(3)$   & $-0.05(5)$            & \checkmark   \\ 
                    RLTSP, $d = 1$              & $-0.336(12)$ & $-0.004(13)$          & \checkmark   \\ 
                    STSP,                       & $-1.5(5)$    & $\phantom{-}0.767(1)$ & degenerate   \\ 
                    (1, 2)-TSP, $\epsilon = 10$ & $-0.82(2)$   & $\phantom{-}0.636(2)$ & RSB possible \\ 
                    (1, 2)-TSP, $\epsilon = 20$ & $-0.93(3)$   & $\phantom{-}0.645(2)$ & RSB possible \\ 
                    (1, 2)-TSP, $\epsilon = 30$ & $-0.97(4)$   & $\phantom{-}0.648(3)$ & RSB possible \\ 
                    $c_{ij} = 1$                &              & $\phantom{-}1$        & degenerate   \\
                \end{tabular}
            \end{ruledtabular}
        \end{table}

    \section{Conclusion}
        To summarize, we studied multiple ensembles of the TSP by applying
        sophisticated exact combinatorial optimization algorithms in extensive
        simulations. As suspected before, we find evidence for the replica
        symmetry of the Euclidean TSP and the related random link model.
        Interestingly, we find this results also in high space dimensions,
        in contrast to spin glasses where RSB is believed to appear above the upper
        critical dimension $d_\text u=6$. Our results strengthen the conjecture
        that replica symmetry holds for these ensembles, which is
        often used to tackle this problem from a statistical mechanics point of
        view.

        On the other hand, for the (1,2)-TSP, inspired by the classical mapping
        of the Hamilton circuit to the TSP, we can not exclude replica symmetry
        breaking. Thus, we provide the first evidence for a complex phase-space
        behavior of this classical NP-hard optimization problem. This should
        motivate further studies to find out whether the solution space is
        clustered and whether replica symmetry breaking might actually be present.

        For future work, especially for the degenerate case of the (1,2)-TSP it
        would be interesting to study the solution space
        structure with a focus on clustering. One could define a
        neighborhood relationship in the configuration space, e.g., $k$-opt
        moves~\cite{lin1965computer}, and search
        for clusters of configurations which can be reached from each other by
        paths traversing only neighboring instances~\cite{hartmann2001ground,Barthel2004Clustering,hartmann2008solution,Montanari2008Clusters}.

        The linear programming approach we use is very general and can be
        applied to a large range of problems. Since for many problems mappings
        to integer programs are already known and it is quite straight forward
        to formulate additional constraints enforcing some specific excitations,
        this technique could be quite generally used to explore a very specific
        range of the energy landscape of many problems.

    \section*{Acknowledgments}
        We thank A.~P.~Young for insightful discussions.
        JKJ thanks the German Academic Exchange Service (DAAD) and the
        International Association for the Exchange of Students for Technical
        Experience (IAESTE) for enabling the research visit to Oldenburg.
        HS thanks the German Research Foundation (DFG) for the grant
        HA 3169/8-1. The simulations were performed at the HPC cluster of the
        GWDG in G\"ottingen (Germany) and CARL, located at the University of
        Oldenburg (Germany) and funded by the DFG through its Major Research
        Instrumentation Programme (INST 184/157-1 FUGG) and the Ministry of
        Science and Culture (MWK) of the Lower Saxony State.

    \bibliography{lit}

\end{document}